\documentclass[twocolumn,article,amsmath,amssymb,aps]{revtex4}
\usepackage{graphicx}
\usepackage{epsfig}

\begin{document}
\draft
%\preprint{HEP/123-qed}
\title{Sum rule for a Schiff-like dipole moment}
\author{A. A. Raduta$^{a,b,c)}$ and R. Budaca $^{b)}$}

\address{$^{a)}$Department of Theoretical Physics,
Bucharest University, POB MG11, Romania}
\address{$^{b)}$Institute of Physics and Nuclear Engineering, Bucharest,
POB MG6, Romania}
\address{$^{c)}$Academy of Romanian Scientists, 54 Splaiul Independentei, Bucharest 050094, Romania}
\date{\today}
\begin{abstract}
The energy-weighted sum rule for an electric dipole transition operator of a Schiff type differs from the Thomas-Reiche-Kuhn sum rule by several corrective terms which depend on the number of system components, ${\cal N}$. For illustration the formalism was applied to the case of $Na$ clusters. One concludes that the RPA results for  $Na$ clusters obey the modified TRK sum rule. 
\end{abstract}
\narrowtext
\maketitle

\renewcommand{\theequation}{1.\arabic{equation}}
\setcounter{equation}{0}
\section{Introduction}
The Thomas-Reiche-Kuhn (TRK) sum rule \cite{Thomas,Kuhn,Reiche} has  been widely used in various contexts of electron excitations in atoms,
molecules, and solids. Indeed,  the  relation to the photoabsorbtion oscillator strength \cite{Orlandini,Au,Alasia,Burton} makes it quite useful in interpreting various collective features of the dipole states \cite{Broglia,Reinhard,Rabura}.

The many body properties like collective excitations in atoms, molecules or clusters are studied using different collective models. Thus, the random phase approximation (RPA) formalism has been extensively used and  excellent results for the photoabsorbtion cross section spectra, especially in the atomic clusters domain \cite{Broglia,Reinhard,Rabura} were obtained.

In a previous publication \cite{Rabura}, some of the many body features of the small and medium sodium clusters  were studied within the RPA approach using the projected spherical single-particle basis defined in Ref.\cite{RaRaRa}. The RPA wave functions were used to treat the dipole transitions which led to the photoabsorbtion cross section spectra. Also,  the system static electric polarizability was analytically determined. 

The salient feature of the RPA approach with spherical single particle basis consists of that it satisfies the TRK sum rule. This is however true for the electric dipole moment, which is not the case in the above quoted paper where, indeed,  a modified dipole operator, similar with the Schiff dipole moment \cite{Schiff,Flambaum,Flambaum1,Zelevinsky1,Zelevinsky2} is used. 
Using the standard dipole operator our formalism is not able to describe the dipole excitations beyond an energy of about $3 eV$,  because one cannot account for the transitions between states characterized by $\Delta N =3$. 

The question which, however, should be answered is whether the use of the Schiff momentum is imposed by a physical constraint or by a technical difficulty. We attempt to provide the answer below. The dipole moment manifests itself, for example, in the interaction of atoms with external electric field which results in
inducing a Stark energy shift. A two body dipole interaction of dipole-dipole type violates the conservation of the center of mass momentum. Treating the interacting system of electrons moving in the electrostatic potential created by the ionic core, and an external electric field in the reference frame where the total momentum of the electron system is conserved one obtains two important results: 1) The total electric field acting on the ionic core is equal to zero, this result being known under the name of the {\it Schiff theorem}. 2) The interaction of electrons with the core is described by the electrostatic potential, generated by the ionic core, and a screening term which is proportional to the gradient of the first. The total potential energy can be written as a series of the average values $<r^n>$. The first non-vanishing $P,T$-odd term of this expansion is a mixture of the octupole and the Schiff momenta. Therefore, the correction to the dipole appear to be a recoil term which assures the restoration of the conservation law for the center of mass linear momentum. 
In conclusion, the correction to the dipole operator accounts for the screening effect, caused by the electronic shells, on the motion of the valence electrons in the mean field determined by the ionic charge.  Moreover the approach of Ref.\cite{Rabura} uses a projected spherical single particle basis which allows for an unified description of spherical and deformed clusters.  

Here we address the question whether the specific features  of our approach require a modification of the TRK sum rule. {\it The result of our investigation is that  a new sum rule holds when one uses the Schiff-like dipole momentum in order to describe the photoabsorbtion spectra in $Na$ clusters.}

In what follows we shall show how did we arrive at this  interesting result. First we provide few basic information.

\renewcommand{\theequation}{2.\arabic{equation}}
\setcounter{equation}{0}

\section{The energy weighted sum rule}
Within the RPA formalism, for any Hermitian operator $\hat{M}$,  the following sum rule holds
\begin{equation}
\sum_{n}(E_{n}-E_{0})\left|\langle0|\hat{M}|1_{n}\rangle\right|^{2}=\frac{1}{2}\langle0|[[\hat{M},H],\hat{M}]|0\rangle,
\label{doublecom}
\end{equation}
where $E_{n}$ are the RPA energies  associated to the many body Hamiltonian $H$. Here the state $|0\rangle$ is the RPA phonon vacuum, while  $|1_{n}\rangle$ denotes the single phonon state
$|1_n\rangle =C^{\dagger}_n|0\rangle$. 

This result is known under the name of Thouless's theorem \cite{Thoul}. 
The summation in Eq. (\ref{doublecom}) suggests that the sum rule may give information about the global properties of the many body system.
Indeed, by using the the lowest k-th energy weighted moment of the strength function one obtains  simple expressions for the mean energy (or energy centroid) and the variance  of the strength function for the linear response of the system to an excitation operator. Lower and upper bounds for the mentioned quantities can be analytically obtained. Also the static polarizability of a cluster is proportional to the sum rule $S_{-2}$.

We are interested in those properties of atomic clusters which are determined by the motion of the valence electrons. These move in a mean field, determined by the ionic core, and interact among themselves through a Coulomb force. The residual two-body interaction \cite{Ekardt,Brack} can be expanded in multipole series from which only the dipole term is relevant and therefore considered. The phonon operator is defined as:
\begin{equation}
C_{n}^{\dagger}(1,\mu)=\sum_{ph}\left[X_{ph}^{n}(c_{p}^{\dagger}c_{h})_{1\mu}-Y_{ph}^{n}(c_{h}^{\dagger}c_{p})_{1\mu}\right],
\end{equation}
where $X_{ph}^{n}$ and $Y_{ph}^{n}$ are the $n$-th order solution of the RPA equations. $c^{\dagger}_p$ and $c^{\dagger}_h$ denote the creation operators for particles and holes, respectively. 
The reduced probability for the dipole transition $|0\rangle \to |1_{n}^{-}\rangle$ can be written in terms of the RPA phonon amplitudes and the $ph$ matrix elements of the transition operator $^{[1]}$ \setcounter{footnote}{1}\footnotetext{ Throughout this paper the Rose's convention for the reduced matrix elements are used.}: 
\begin{equation}
B(E1,0^{+}\rightarrow1_{n}^{-})=\left|\langle0||{\cal M}(E1)||1_{n}^{-}\rangle\right|^{2}.
\end{equation}
Instead of the usual transition dipole operator, a Schiff-like moment operator \cite{Schiff,Zelevinsky1,Zelevinsky2} was used.
\begin{equation}
\mathcal{M}(E1)=e\left(1-\frac{3}{5}\frac{r^{2}}{r_{s}^{2}}\right)\vec{r}.
\label{Schiff}
\end{equation}
Here $r_{s}$ is the Wigner-Seitz radius  which, for $Na$ clusters has the value of 3.93 a.u.. The corrective component, involved in the dipole operator, may relate particle and hole states characterized by $\Delta N =3$, which results in modifying the strength distribution among the RPA states. 
Such an effect would be however obtained even for the dipole transition operator, if the mean field potential for the single particle motion involves higher powers of the radial coordinate.

Reckoning the double commutator from Eq. (\ref{doublecom}), corresponding to the transition operator (\ref{Schiff}) one obtains:

\begin{eqnarray}
&&\sum_{n}(E_{n}-E_{0})\left|\langle0||\mathcal{M}(E1)||1_{n}\rangle\right|^{2}=\frac{9\hbar^{2} e^{2}}{2m_{e}}\nonumber\\
&&\times\left[\mathcal{N}
-\frac{2}{r_{s}^{2}}\langle 0||\sum_{\alpha=1}^{\mathcal{N}}r_{\alpha}^{2}||0\rangle+
\frac{33}{25r_{s}^{4}}\langle 0||\sum_{\alpha=1}^{\mathcal{N}}r_{\alpha}^{4}||0\rangle\right].
\label{SR1}
\end{eqnarray}
 
 The  terms  correcting the standard TRK sum rule are the expected values of the radius powers $r^2$ and $r^4$ , in the RPA ground state. 
Obviously, these corrective terms induce an additional ${\cal N}$ dependence for the 
energy-weighted sum of the reduced dipole transition probabilities. In what follows we shall use the notation $EWS$ for the left side  and   $S({\cal N})$ for the right hand side of Eq. (\ref{SR1}). $EWS$  can be directly calculated using the RPA output data, like energies and transition probabilities \cite{Rabura}.
{\it If $S({\cal N})$ has a model independent expression involving only universal constants, and the equality $EWS=S({\cal N})$ holds, one says that a sum rule is valid for the system under consideration.} Since $EWS$ is calculated by complicated many body formalisms, obeying the sum rule equation is a serious test for the proposed description. 
The terms of $S(\mathcal{N})$ were alternatively evaluated through two distinct methods:
\subsection{ The boson expansion method}
The terms involved in $S(\mathcal{N})$ can be expressed in terms of particle-particle ($pp$) and hole-hole ($hh$) transition matrix elements. Indeed, the $ph$ transition components give vanishing contributions when they are averaged with the RPA ground state. Therefore, the needed one-body operator $\hat{r}^m$ can be written as:  
\begin{equation}
\sum_{\alpha=1}^{\mathcal{N}}r_{\alpha}^{m}\equiv\sum_{p}\langle p|r^{m}|p\rangle c_{p}^{\dagger}c_{p}+\sum_{h}\langle h|r^{m}|h\rangle c_{h}^{\dagger}c_{h}.
\end{equation}
The fermion density operators $c_{p}^{\dagger}c_{p}$ and
 $c_{h}^{\dagger}c_{h}$ can be expressed  in terms of the RPA phonon operators $C_{n}^{\dagger}(C_{n})$: 
\begin{equation}
c_{p}^{\dagger}c_{p}=\sum_{n}a_{p}^{n}C_{n}C_{n}^{\dagger},\;
c_{h}^{\dagger}c_{h}=\sum_{n}b_{h}^{n}C_{n}C_{n}^{\dagger},
\end{equation}
where the coefficients $a_{p}^{n}$ and $b_{h}^{n}$ have the expressions:
\begin{equation}
a_{p}^{n}=\langle 0|[[C_{n}^{\dagger},c_{p}^{\dagger}c_{p}],C_{n}]|0 \rangle,\;
b_{h}^{n}=\langle 0 |[[C_{n}^{\dagger},c_{h}^{\dagger}c_{h}],C_{n}]|0 \rangle.
\end{equation}

In this way the modified dipole sum rule becomes:
\begin{eqnarray}
&&\sum_{n}(E_{n}-E_{0})\left|\langle 0||\mathcal{M}(E1)||1_{n}\rangle\right|^{2}=\frac{9\hbar^{2} e^{2}}{2m_{e}}\left[\mathcal{N}+\right.\\
&&\left.\sum_{n,p,h}\left((X^{(n)}_{ph})^2+(Y^{(n)}_{ph})^2\right)\left[-\frac{2}{r_s^2}r^2(ph)+\frac{33}{25r_s^4}r^4(ph)
\right]\right],\nonumber
\label{MSR}
\end{eqnarray}
where the following notation has been used:
\begin{equation}
r^k(ph)=\langle p|r^k|p\rangle -\langle h|r^k|h\rangle ,\,k=2,4.
\end{equation} 
As we have already mentioned, a specific ingredient of the  approach from Ref. \cite{Rabura} is the use of the projected spherical single particle basis in the RPA formalism.

\subsection{ Electron density approach}
Static electric polarizabilities results based on the calculations for the number of spilled out electrons \cite{Rabura}, agree quite well with the corresponding experimental data. These spilled out electrons produce a screening effect against external fields which results in changing the classical result for the polarizability. The basic assumption in accounting the spilled out electrons is the fact that the electron density is not going sharply to zero at the cluster surface, but is gradually decreasing and moreover extends significantly beyond the jellium edge. The same argument can be brought for the correction terms of the $S(\mathcal{N})$ containing averages of the radius powers
with the RPA vacuum state. The electronic density has a constant central part, enfolded by a diffuse region, of a width equal to $a$, where the electron density tends smoothly to zero. Guided on some formal parallelism between the behaviors of atomic clusters and nuclear systems \cite{Koskinen,Rigo}, the average of $r^m$ with the RPA vacuum state is approximated by folding $r^m$ with a localization probability density with spherical symmetry, of Fermi distribution type.
\begin{equation}
\rho(r)=\rho_0\left[1+\exp\left(\frac{r-R}{a}\right)\right]^{-1}.
\end{equation}
Here $R=r_{s}\mathcal{N}^{1/3}$ is the radius of the cluster with $\mathcal{N}$ atoms.
$a$ is a parameter defining the thickness of the diffusion region. For nuclear systems, $a$ is a constant related to the diffuseness parameter $d$ by a simple relation: $d\approx 4.4a.$

For atomic clusters such a density function was used, in a different context, by Snider and Sorbello \cite{Snider}. 
A modified density function, having the denominator at a power $\gamma\ne 1$,  which yields a density profile with an asymmetry around the inflexion point, was used in Ref.\cite{Brack}.

Having the density, momenta $<r^m>$ may be written in the form of a power series in the variable $\frac{a}{R}$ \cite{Nils}:
\begin{eqnarray}
\langle r^{m}\rangle &=&R^{m}\frac{3}{m+3}\left[1+\frac{\pi^{2}}{6}\left(\frac{a}{R}\right)^{2}m(m+5)+\ldots\right],\nonumber\\
&&m=2,4.
\end{eqnarray}
  
In this way the term $S(\mathcal{N})$ can be written in the following way
\begin{eqnarray}
&&S(\mathcal{N})=\frac{9\hbar^{2}e^{2}}{2m}\left[\mathcal{N}-\frac{6}{5}\mathcal{N}^{2/3}+\frac{99}{175}\mathcal{N}^{4/3}+\right.\nonumber\\
&&\left.\frac{\pi^{2}a^{2}}{5r_{s}^{2}}\left(\frac{594}{35}\mathcal{N}^{2/3}-14\right)\right]
\equiv \frac{9\hbar^{2}e^{2}}{2m}{\cal F(N)}.
\label{SofNanda}
\end{eqnarray}
One notices that besides the number of atoms dependency, this expression involves only universal constants, which makes Eq. (\ref{SofNanda}) to be, indeed, a real sum rule. 

Contrary to the case of nuclear systems, where the thickness of the diffusion region is approximately the same for all nuclei,  for atomic clusters the parameter $a$  is expected to have a ${\cal N}$ dependency due to the long range character of the two body interaction. This dependence was determined by interpolating the values of $a$ satisfying the equation
$EWS =S({\cal N})$, for $8\le {\cal N}\le 40$. The solutions for $a$ were interpolated by the function of ${\cal N}$:

\begin{equation}
a(\mathcal{N})=-0.975-0.011\mathcal{N}^{1/3} +0.361\mathcal{N}^{2/3}.
\label{difus}
\end{equation}
Inserting the expression (\ref{difus}) of $a(\mathcal{N})$  in Eq.(\ref{SofNanda}), one obtains the final expression for the sum $S({\cal N})$. 

The sum rule is always a serious consistency test for any model calculation and in particular for the RPA formalism of Ref.\cite{Rabura} which uses a  projected spherical single particle basis, appropriate for an unified description of spherical and deformed clusters, and a Schiff-like momentum as a dipole transition operator. 

Since the double commutator appearing in the sum rule (2.1) is averaged with the ground state, it results that the sum rule reflects the structure of the ground state.

We would like to mention that Eq.(\ref{doublecom}) is valid also when the sates involved are exact eigenstates while  energies are the
exact eigenvalues of H \cite{Sanwu}. Note that despite the fact RPA is an approximative approach the TRK sum rule is exactly obeyed. Due to this feature it is a legitimate question {\it which are the approximations which preserve the sum rule} and which are those which violate it. Of course this question cannot be completely answered. Despite of the fact that this issue is more suited for a revue paper rather than for
a short letter, here we give some brief comments on this matter. 

If we replace the RPA dipole  states by the particle-hole dipole states 
the TRK sum rule is still satisfied. In nuclear physics the N-Z sum rule for the Gamow-Teller transition operator is also satisfied in the framework of BCS approximation. In this case the proton-neutron quasiparticle  RPA dipole states are replaced by two proton-neutron quasiparticle dipole states. Aiming at satisfying the Pauli principle, the pnQRPA has been improved by a renormalization procedure \cite{Suho,Rad98}, while the transition operator was left unmodified. With these new ingredients the $N-Z$ sum rule is  violated by about 25-30\%. Another approach for this transition process was formulated by one of the present authors (A.A.R.) \cite{Rad91}, by taking for the transition operator a first order boson expansion expression while the states involved for the mother and daughter nuclei were still described within the pnQRPA formalism. Again 
the deviation from the sum rule is about 20-25\%. It is interesting to remark that the first procedure underestimates the sum rule while the second one overestimate it. This suggested us to use for the GT transition operator a boson expansion expression in terms of the pnQRPA renormalized boson. The states assigned for the nuclei participating to transitions are described by the renormalized pnQRPA formalism.
In this way the deviation of the sum rule from the $N-Z$ value was diminished up to less than 10\%.

An inconsistency which might induce a deviation from the TRK sum will be discussee a bit latter,   when the numerical results
for the boson expansion method applied to the expected values of $r^2$ and $r^4$, are presented.

The structure of the mean field may be a source of inconsistency. Indeed, if the mean field is of a harmonic oscillator type the dipole operator is not able to connect states characterized by $\Delta N=3$ and consequently the volume like collective excitation would be missing.

An important feature related to the sum rule is the completeness of the states defined within the chosen approximation. For example, the RPA calculations
  are performed within a restricted single particle  and hole space which determine the unity resolution for the RPA space(
$1=\sum_{k}|1_k\rangle\langle 1_k|$) which, in its turn, yields a certain value of the sum rule. If we enlarge the single particle space,  the new RPA space determines a new unity resolution and consequently a new value for the sum rule. In general, the sum rule depends on the dimension of the single particle space used by the RPA approach. In principle the sum rule is an increasing function of the dimension of the particle-hole configurations. How fast the sum rule converges when this dimension is increased depends of course on the single particle basis provided by the employed mean field as well as on the transition operator.

The exchange term involved in the electron two body interaction yields a screening effect for the system response to the action of an exciting external field. However such a screening effect is not found in the transition dipole operator. As a matter of fact  this is the
 inconsistency we wanted to remove when we used the Schiff-like  dipole operator. Moreover, such an operator links the $\Delta N=3$ states and, therefore, yields a non-vanishing strength for the volume dipole excitation.  

In our formalism the RPA procedure uses a projected spherical single particle basis and a Schiff-like dipole operator. Due to these new ingredients we expect a new sum rule. As we have seen before, the corresponding EWS has a simple expression as a function of ${\cal N}$ and consequently satisfies the commonly used definition for a sum rule. The consistency of our calculations is tested by checking numerically whether the equation \ref{MSR} is satisfied.

\renewcommand{\theequation}{3.\arabic{equation}}
\setcounter{equation}{0}

\section{Numerical application}

The $EWS$ was calculated with the RPA energies and matrix elements for the transition operator. $S({\cal N})$ was alternatively calculated with the expressions  (9) and (13), respectively. The two sets of results corresponding to the mentioned options for $S({\cal N})$, are plotted in Figs. 1 and 2, respectively.

The RPA calculation of $S({\cal N})$ does not yield an explicit ${\cal N}$ dependence although such a dependence is involved in the many body formalism by means of the Fermi energy, the oscillator length, the matrix elements in the projected single particle basis and the space of the single particle states used in the RPA calculation. Note that  Fig. 1 shows a good agreement for medium clusters. The deviation is significant for large as well as for small clusters although both curves exhibit similar pattern concerning the oscillating behavior. The discrepancies may indicate that higher order boson expansion terms are necessary in order to improve the agreement. However, these terms would bring a certain inconsistency to the formalism since $EWS$
is evaluated within the RPA approach. Moreover, in order that Eq.(9) plays the role of a sum rule it is necessary that  $S({\cal N})$ exhibits a model independent expression.  

In this context it is worth mentioning that the second procedure described above, yields indeed, a model independent expression for $S({\cal N})$. The results for this situation are shown in Fig. 2 where
 a very good agreement between $S({\cal N})$ and  $EWS$ is shown. 

\begin{figure}[hbtp!]
\begin{center}
\includegraphics[width=0.45\textwidth]{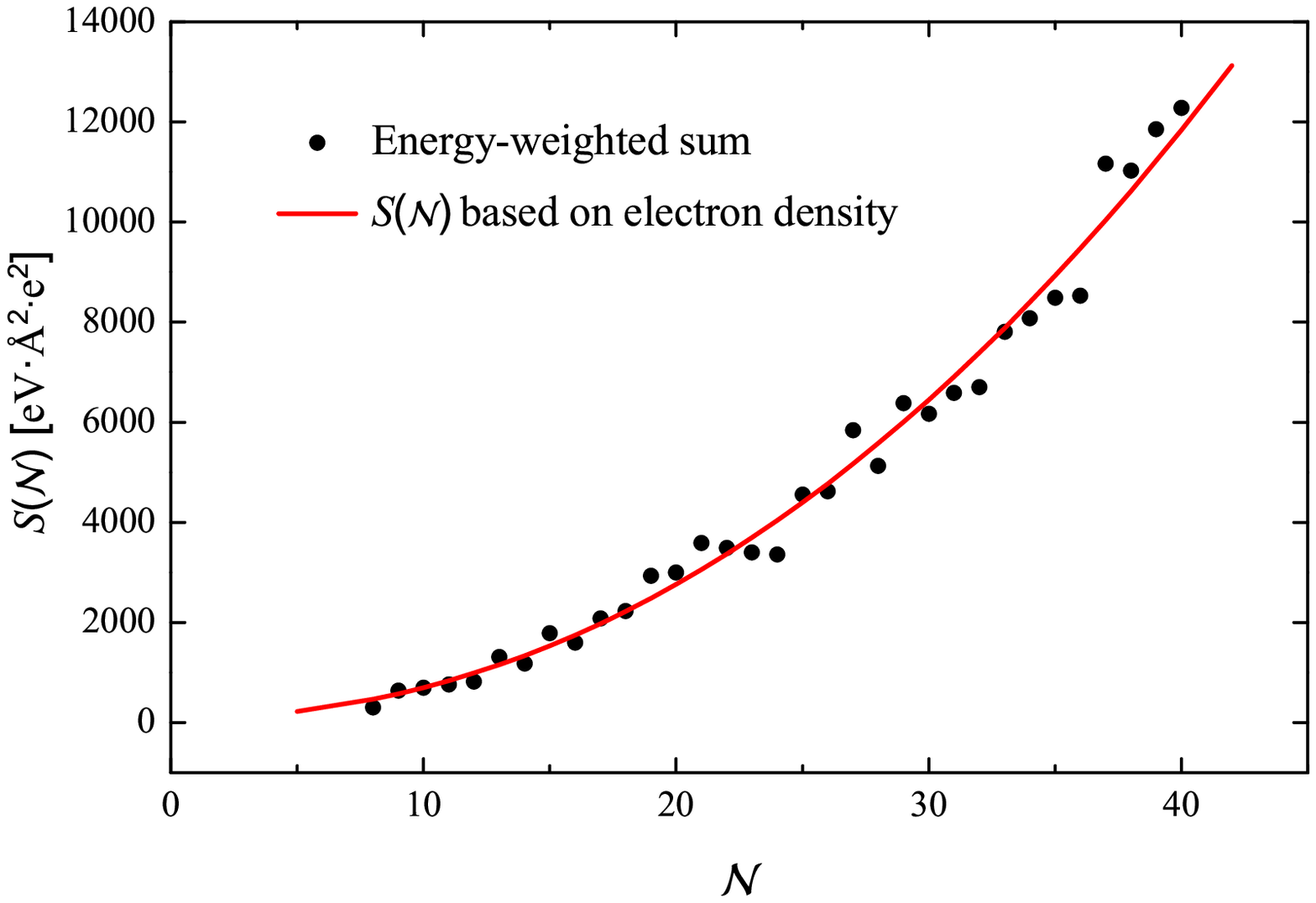}
\end{center}
\vspace*{-0.5cm}
\caption{The calculated $EWS$ (open circles)  and $S(\mathcal{N})$ given by the RPA approach (black triangles),
versus ${\cal N}$, the number of cluster's components.}
\end{figure}

\begin{figure}[hbtp!]
\begin{center}
\includegraphics[width=0.45\textwidth]{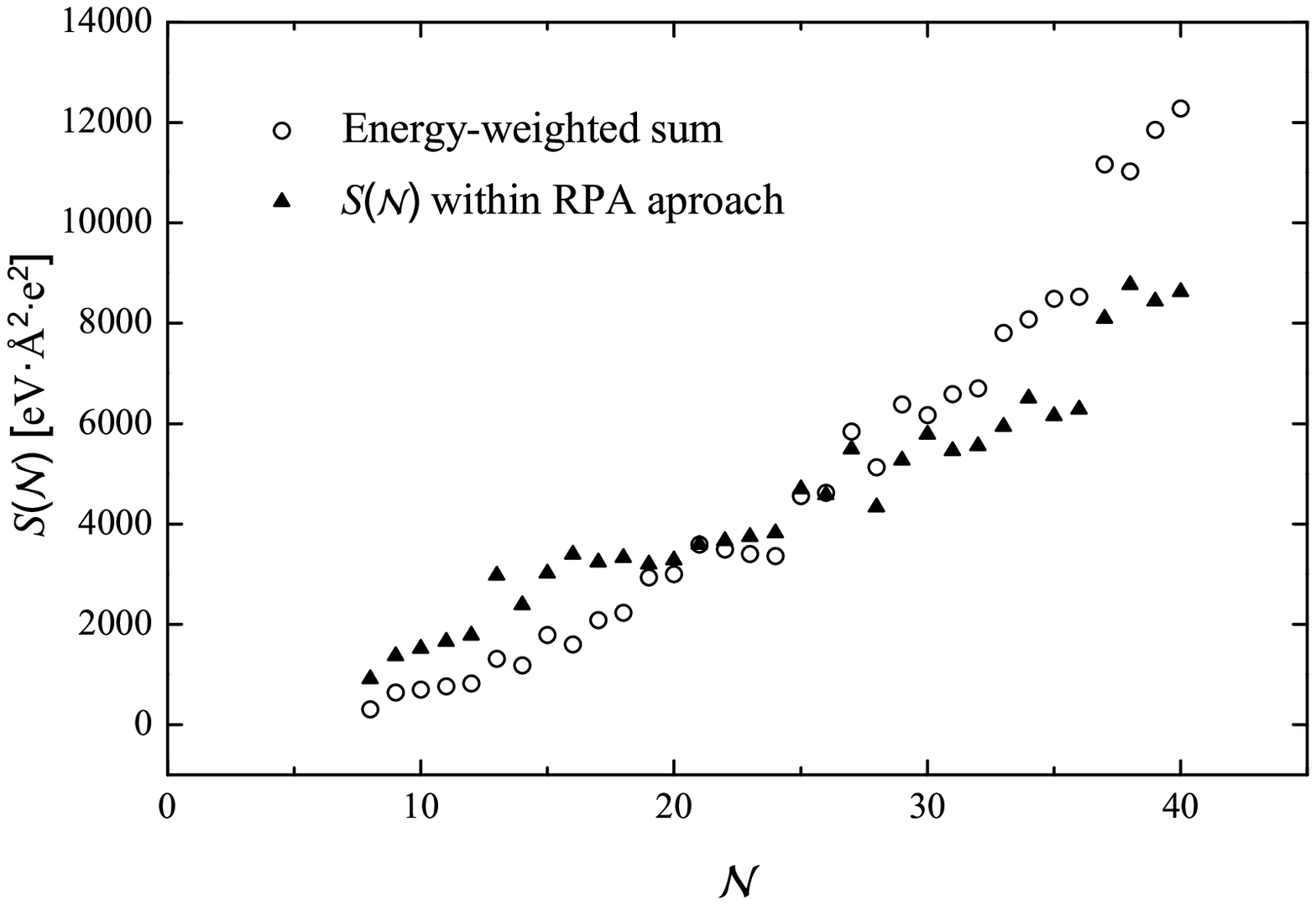}
\end{center}
\vspace*{-1cm}
\caption{The same as in Fig. 1 but $S(\mathcal{N})$  given by Eq. (13) and represented by a solid line. Theoretical result
a given by black circles.}
\end{figure}

{\it Actually this is the main result of our investigation. The sum rule for the modified dipole operator involves $S({\cal N})$ given by Eq.(2.13) with the  parameter $a$ from (2.14).}  By contrast to  the TRK sum rule, the new one is not linear in ${\cal N}$.

Of course one may ask how the sum rule is mirrored by the experimental results. Direct data for the momenta $<r^2>$ and $<r^4>$ are not yet available. Under these circumstances we have  to see
 whether the experimental data for the $EWS$ agree with the calculated values.
 Actually, the experimental value for $EWS$ can be extracted from the experimental photoabsorbtion cross sections. Indeed, interpolating the discrete values of the experimental photoabsorbtion cross sections, given as function of the excitation energy, by a smooth curve and integrating the result with respect to the energy in the interval $[0,\infty)$, one obtains the area ${\cal A(N)}$. The experimental $EWS$ is proportional to ${\cal A(N)}$. We assume that the ${\cal N}$ depending proportionality factor is  the same as for the theoretical
$EWS$, i.e. ${\cal F(N)}$. This assumption is grounded  on the fact that  ${\cal F(N)}$ is a model independent quantity and moreover assures similar normalization for the total cross section as in the schematic calculation. Thus, the experimental $EWS$ is defined by:
\begin{equation}
\left[EWS\right]_{Exp.}={\cal GF(N)A(N)}.
\end{equation}
The quantity ${\cal A(N)}$, extracted from the data of Refs.\cite{Voll,Selby1}, varies between
0.582 (eV){\AA}$^2$ for ${\cal N} =9$ and 0.387 (eV){\AA}$^2$, for ${\cal N}=19$.
Here the constant factor is ${\cal G}=70.159 [e^2]$. This value was obtained by equating the right hand side of the above equation to the calculated value for $EWS$ for the case of $Na_{14}$, were the best agreement between the calculated and experimental polarizabilities was obtained.
Thus, the constant ${\cal G}$ yields a normalization of $EWS$ which accounts for the "missing sum rule", noticed experimentally \cite{VitalKre}. 

Actually, this normalization of the integrated cross section might be used  for certifying the collective character of a dipole excitation, according to its contribution to the sum rule. Thus the collective surface dipole excitations exhaust between 70\% to 100\% from the sum rule. In the case of small and medium clusters the remainder sum rule is attributed either  to a volume-like excitation \cite{Kresin} or to  single electron excitations \cite{Selby1}.  

\vspace*{-0.2cm}
\begin{figure}[hbtp!]
\begin{center}
\includegraphics[width=0.45\textwidth]{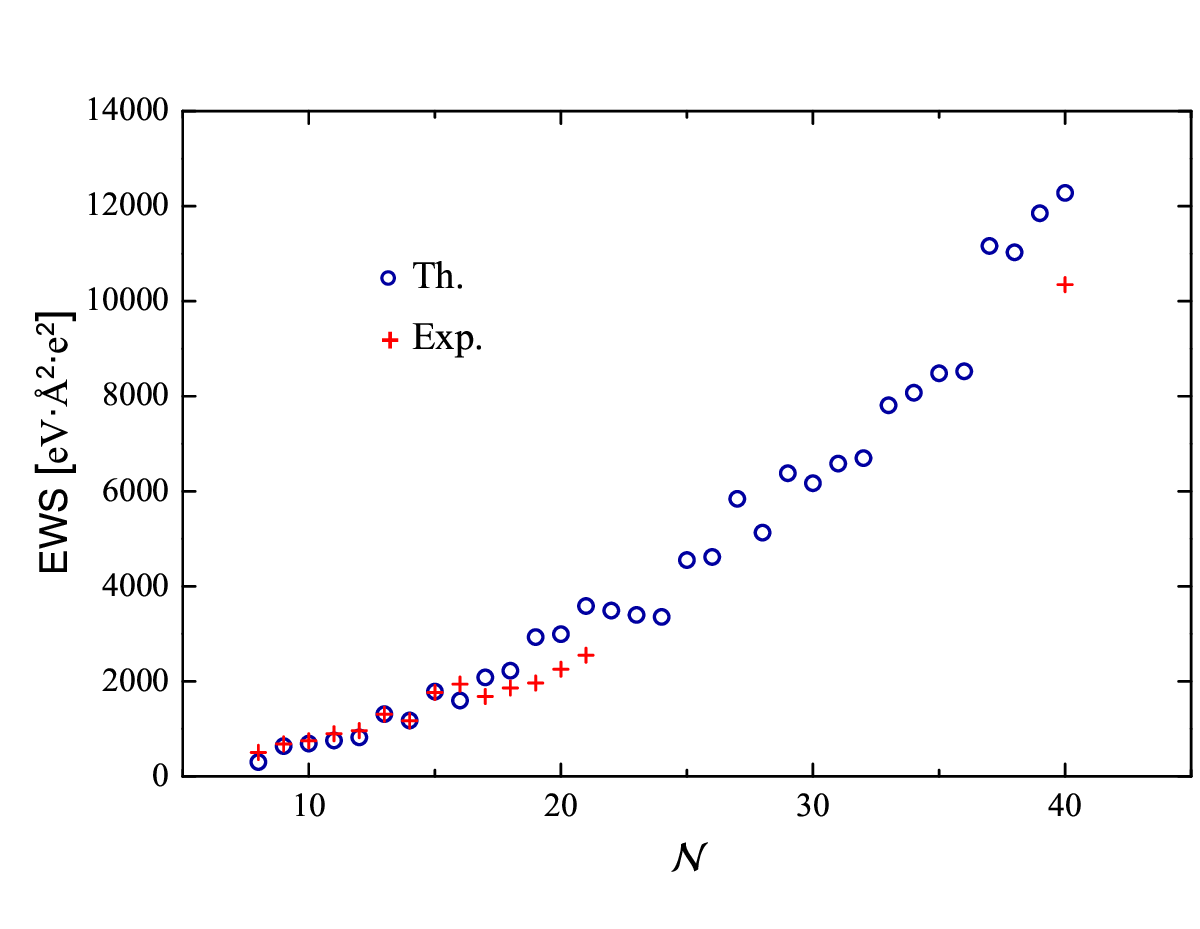}
\end{center}
\vspace*{-1cm}
\caption{The quantities$[EWS]_{Exp.}$ (crosses) and  $[EWS]_{Th.}$ (open circles) defined by Eqs. (15) and (16) respectively, are given as function of the number of cluster's components, ${\cal N}$.}
\end{figure}

In Fig.3, the quantity $\left[EWS\right]_{Exp.}$ is  compared  with the theoretical weighted sum $EWS$, calculated within the RPA approach, which uses a projected spherical single particle basis \cite{Rabura},
\begin{equation}
\left[EWS\right]_{Th.}=\left[\sum_{n}(E_n-E_0)B(0^{+}\to 1^{-}_n)\right]_{RPA}.
\end{equation} 
The agreement between the two $EWS$'s, shown in Fig. 3, is a guarantee that $\left[EWS\right]_{Exp.}$ is, indeed, close to the ${\cal S(N)}$. 
Note that in Fig.3 we didn't consider  clusters with $21<{\cal N}<40$ although for some of them experimental data are available. The reason is that for these data only few cross sections are given and therefore it is not possible to calculate the area ${\cal A}$ introduced above.

\renewcommand{\theequation}{1.\arabic{equation}}
\setcounter{equation}{0}

\section{Conclusions}

Finally, we want to mention that the octupole correction to the dipole transition operator was also used 
in connection with the description of the electric dipole transitions in nuclear systems. Thus, in Ref.\cite{Ham} it is pointed out that adding the octupole correction,  the agreement with experimental data concerning the
E1 transitions is substantially improved. In Ref.\cite{Solo} a similar effect is obtained by modifying the many body wave function due to the octupole interaction which is considered in addition to an isovector-dipole interaction. 
The new components are connected by the standard dipole operator and consequently modifies the E1 transition rates. Of course,
an energy-weighted sum rule associated to the dipole transition operator holds also for nuclear systems. Obviously, changing the transition operator, as it happened in Ref.\cite{Ham}, the corresponding sum rule which should be valid is the one obtained in the present paper.

Another interesting example is the $N-Z$ sum rule, which holds for the single beta transition. This rule says that for a single beta decaying nucleus, the difference between the $\beta^-$  and $\beta^+$ strengths should be equal to $3(N-Z)$. One remarks that this is in fact the nuclear physics counterpart of the TRK sum rule, which may be alternatively formulated  as expressing the equality of the difference between the photoabsorbtion  and emission dipole transition probabilities and the number of delocalized electrons. The particle-hole RPA approach satisfies exactly the TRK sum rule.
 The N-Z sum rule is true, for example, for Gamow-Teller (GT) dipole transitions and is exactly satisfied within the proton-neutron quasiparticle RPA approach. Extensions of the microscopic formalisms to the double beta decay $2\nu\beta\beta$, showed that in order to describe the transition rates, it is necessary
to improve the wave functions of the mother nucleus as well as the GT dipole states, by adding anharmonic effects. However, these corrections violate drastically the $N-Z$ sum rule. In the spirit of the present paper, we open the question whether the Gamow-Teller proton-neutron interaction could be extended by adding an octupole component such that to the new proton-neutron interaction a modified $N-Z$ sum rule corresponds. This would make the inclusion of anharmonic effects which, as a matter of fact, violates the Pauli principle, unnecessary.
 It is an open question whether a Schiff like correction of the Gamow-Teller transition operator is necessary due to some specific conservation law, or just due to the necessity of improving the existent descriptions.

The final conclusion is  that the Schiff-like dipole moment used for the RPA description of the photoabsorbtion cross section spectrum, satisfies an extended TRK  sum rule. The saturation of the extended sum rule is a positive test for the single particle basis as well as for the dimension of the dipole $ph$ space involved in the RPA description. The  TRK sum rule is of a general interest, being applicable also for other many body systems correlated by a Schiff-like two body interaction. The usefulness of sum rules in exploring the many body properties mirrored by the multipole electric, or magnetic transitions have been stressed by many authors  \cite{Bohig,Kresin,Moya}. Here we showed that a sum rule may hold also for a multipole mixed transition operator. 

\noindent
{\bf Acknowledgments}
This work was also supported  by the Romanian Ministry for Education and Research under the contracts PNII ID-33/2007 and ID-946/2007 .

\end{document}